\documentclass{article}
\usepackage{amsmath}
\usepackage{amsthm}

\usepackage{enumerate} 
\usepackage{upgreek} 
\usepackage{color} 
\usepackage{xcolor} 
\usepackage{mathtools} 
\usepackage{tikz}
\usetikzlibrary{matrix,graphs,arrows,positioning,calc,decorations.markings,shapes.symbols,shapes.geometric,shapes.misc}
\usepackage{calc, ifthen, ae, xstring, etoolbox, xkeyval,pgfplots}
 \usepackage{float}
 
 \usepackage[english]{babel}
 \usepackage[pdftex,bookmarks,colorlinks,breaklinks]{hyperref}  
 \addto\extrasenglish{
}
 \addto\extrasenglish{
}

\newtheorem{theorem}{Theorem}[section]
\newtheorem{proposition}[theorem]{Proposition}

\theoremstyle{remark}
\newtheorem{remark}[theorem]{Remark}

\newcommand{\h}{\operatorname{H}}

\newcommand{\pain}[1]{\text{P}_{\mathrm{#1}}}

\newcommand{\Ok}{\operatorname{Ok}}
\newcommand{\KNY}{\operatorname{KNY}}
\newcommand{\FZ}{\operatorname{FZ}}
\newcommand{\IP}{\operatorname{IP}}

\numberwithin{equation}{section}

\begin{document}
\title{Relations between different Hamiltonian forms of the third Painlev\'e equation} 
\author{
Galina Filipuk\footnote{Institute of Mathematics, University of Warsaw, ul. Banacha 2, 02-097, 
Warsaw, Poland.
Email: filipuk@mimuw.edu.pl}, 
Adam Lig\c{e}za\footnote{Institute of Mathematics, University of Warsaw, ul. Banacha 2, 02-097, 
Warsaw, Poland.
Email: a.ligeza@mimuw.edu.pl}, 
Alexander Stokes\footnote{Graduate School of Mathematical Sciences, The University of Tokyo, 3-8-1 Komaba Meguro-ku Tokyo 153-8914, Japan.
Email: alexander.stokes.14@ucl.ac.uk}}
\maketitle

\begin{abstract}
In this paper we study different Hamiltonian systems with polynomial and rational Hamiltonians associated with the generic third Painlev\'e equation and present explicit birational transformations relating them.
\end{abstract}



\section{Introduction}

A standard form of the third Painlev\'{e} equation is 
\begin{equation}
\pain{III}:\quad \frac{d^2x}{dt^2} = \frac{1}{x}\left(\frac{dx}{dt}\right)^2-\frac{1}{t}\frac{dx}{dt}+\frac{1}{t}(\alpha x^2+\beta)+\gamma x^3+\frac{\delta}{x},
\end{equation}
where $x$ is a function of $t$ and $\alpha,\,\beta,\,\gamma,\,\delta$ with $\gamma\delta\neq 0$ are arbitrary complex parameters.  
There exist in the literature multiple non-autonomous Hamiltonian systems equivalent to $\pain{III}$ obtained from various viewpoints, but the relations between these have not been fully established.
In this paper we continue the program announced in \cite{P4paper}, which aims to compare the different Hamiltonian systems associated with each Painlev\'e equation and find explicit birational (or algebraic) transformations between them. 
We draw on a variety of methods, including the comparison of linear systems from the isomonodromy approach as well as the method presented in  \cite{P4paper}, which uses tools from the geometric theory of Painlev\'e equations developed by Okamoto and Sakai \cite{Ok1, Sak01}.
Obtaining such relations between different Hamiltonian systems related to Painlev\'e equations is sometimes quite subtle, for example for the fourth Painlev\'e equation studied in \cite{P4paper} the relations are very much non-trivial and the geometric method is more effective than a direct approach.
In this paper we consider the case of the third Painlev\'e equation, more specifically the generic case where $\gamma \delta \neq 0$ as above, for which the relations between various Hamiltonian systems turn out to be more straightforward and less interesting on the level of the geometric theory, so we omit details and present only the birational transformations themselves.

The Hamiltonians and the associated systems of differential equations equivalent to $\pain{III}$ that we plan to compare in this paper are the Okamoto (polynomial) Hamiltonian \cite{Ok2, Ok3}, the Kajiwara-Noumi-Yamada (KNY) Hamiltonian \cite{KNY}, the Filipuk-\.Zo\l\c{a}dek rational Hamiltonian \cite{FZ} and two Hamiltonians obtained from the isomonodromic deformation of linear systems: the Jimbo-Miwa Hamiltonian \cite{JM} and the Its-Prokhorov Hamiltonian \cite{IP}. 
We discuss only  the Its-Prokhorov Hamiltonian since up to a function of $t$ and scaling of parameters they are the same.

\section{Main results}


In the papers \cite{Ok2} and \cite{Ok3}, the third Painlev\'e equation appears as the result of the monodromy preserving deformation of a scalar second order linear differential equation (with coefficients that depend on $t$) with certain singularities in the complex plane. The third Painlev\'e equation is obtained in the form of a Hamiltonian system with polynomial Hamiltonian \cite{Ok1, Ok2}, given by
\begin{equation}\label{Ham Ok}
\h^{\Ok} (f,g,t) = \frac{1}{t}(2f^2 g^2+(\theta_0+\theta_{\infty})\eta_{\infty} t f -((1+2\theta_0)f-2\eta_0 t+2\eta_{\infty} t f^2)g),
\end{equation}
where $f=f(t)$ and $g=g(t)$, $t$ is an independent variable, and  $\theta_0, \,\theta_{\infty},\,\eta_0,\,\eta_{\infty}$ are complex constants.
The corresponding Hamiltonian system is
\begin{equation}\label{system Ok}
\left\{ 
\begin{aligned}
\frac{df}{dt} &=\frac{\partial \h^{\Ok} }{\partial{g}}=\frac{ 4f^2 g-2\eta_{\infty}t f^2-(1+2\theta_0)f+2\eta_0t}{t}, \\ 
\frac{dg}{dt} &=-\frac{\partial \h^{\Ok}}{\partial{f}}= - \frac{4f g^2-(1+2\theta_0+4\eta_{\infty}t f)g+\eta_{\infty}(\theta_0+\theta_{\infty})t}{t}.
\end{aligned}
\right.
\end{equation}
Eliminating the function $g$ from the  system above, we obtain $\pain{III}$ for the function $f$ with parameters 
\begin{equation}\label{par Ok}
\alpha= -4\eta_{\infty}\theta_{\infty}, \quad \beta= 4\eta_0(1+\theta_0),\quad \gamma=4\eta_{\infty}^2,\quad \delta=-4\eta_0^2.
\end{equation}

The KNY Hamiltonian \cite[(8.237)]{KNY} is given by 
\begin{equation}\label{Ham KNY}
\h^{\KNY} (q,p,s) = \frac{1}{s}(p(p-1)q^2+(a_1+a_2)q p+s p-a_2 q),
\end{equation}
where $q=q(s)$ and $p=p(s)$, $s$ is an independent variable and  $a_1$ and $a_2$ are complex constants.
The corresponding Hamiltonian system is given by 
\begin{equation}\label{system KNY}
\left\{
\begin{aligned}
\frac{dq}{ds} &=\frac{\partial \h^{\KNY}}{\partial{p}}=\frac{q^2 p+(p-1)q^2+(a_1+a_2)q+s}{s}, \\ 
\frac{dp}{ds} &=-\frac{\partial \h^{\KNY}}{\partial{q}}=-\frac{2(p-1)q p+(a_1+a_2)p-a_2)s}{s}.
\end{aligned}
\right.
\end{equation}
The KNY Hamiltonian system is very convenient to use as a reference example for the geometric approach, since \cite{KNY} provides a comprehensive collection of relevant data. 
However the relation to the third Painlev\'e equation is not as straightforward here as in the Okamoto case. 
According to \cite{KNY}, one needs to eliminate first the variable $p$, then introduce the function $x(t)$ by $q(s)=t x(t)$ with $s=t^2$ and then one obtains the third Painlev\'e equation for $x(t)$ with parameters
\begin{equation}\label{par KNY}
\alpha=-4(a_1-a_2), \quad \beta=4(1-a_1-a_2), \quad \gamma=4, \quad \delta=-4.
\end{equation}

The Filipuk-\.Zo\l\c{a}dek (rational) Hamiltonian \cite{FZ} is given by
\begin{equation} \label{Ham FZ}
\h^{\FZ} (x,y,t) = \frac{x^2 y^2}{2t}-\alpha x+\frac{\beta}{x}-\frac{\gamma t x^2}{2}+\frac{\delta t}{2 x^2},
\end{equation}
where $x=x(t)$ and $y=y(t)$, $t$ is an independent variable and  $\alpha,\,\beta,\,\gamma,\delta$ are complex constants.
The corresponding Hamiltonian system is given by 
\begin{equation}\label{system FZ}
\left\{
\begin{aligned}
\frac{dx}{dt} &=\frac{\partial \h^{\FZ}}{\partial{y}}= \frac{x^2 y}{t}, \\ 
\frac{dy}{dt} &=-\frac{\partial \h^{\FZ}}{\partial{x}}=\alpha- \frac{x y^2}{t}+ \frac{\beta}{x^2}+\gamma t x+\frac{\delta t}{x^3}.
\end{aligned}
\right.
\end{equation}
Eliminating the function $y$ between these equations we immediately get the third Painlev\'e equation for the function $x$, with parameters $\alpha, \beta, \gamma, \delta$.

In \cite{IP}, the authors obtained Hamiltonians related to the Painlev\'e equations from the study of monodromy preserving deformations of $2\times 2$  linear systems, with the $\pain{III}$ case being
\begin{equation}
\h^{\IP}(f,g,t)=\frac{1}{t}(2f^2g^2+(2t-2t f^2+(4 \theta_{\infty}-1)f)g-2t f(\theta_0+\theta_{\infty})+(\theta_{\infty}^2-\theta_0^2)).
\end{equation}
The corresponding Hamiltonian system gives the third Painlev\'e equation for the function $f$, with parameters $\alpha=8\theta_0$, $\beta=4-8\theta_{\infty}$, $\gamma=-\delta=4$. 
For the fourth Painlev\'e equation, the corresponding Its-Prokhorov Hamiltonian was different from that of Okamoto. 
However, in this case we see that if we take Okamoto's Hamiltonian with $\eta_1$ and $\eta_{\infty}$ equal to one, and set $\theta_{\infty}^{\IP}=-\theta_0^{\Ok}/2$, $\theta_0^{\IP}=-\theta_{\infty}^{\Ok}/2$, then up to a function of $t$ both Hamiltonians (and, hence, corresponding systems) coincide.

Hence, we are left with two nontrivial cases of relations between systems to establish: first between the rational Hamiltonian $\h^{\FZ}$ and the Okamoto Hamiltonian $\h^{\Ok}$, and second between the rational Hamiltonian $\h^{\FZ}$ and the KNY Hamiltonian $\h^{\KNY}$. 
In what follows we will present these, completing the picture of how different Hamiltonian forms of $\pain{III}$ are related, as outlined in Figure \ref{fig:allHams}.


\tikzstyle{box} = [rectangle, rounded corners, minimum size=6mm, text centered, draw=black
]

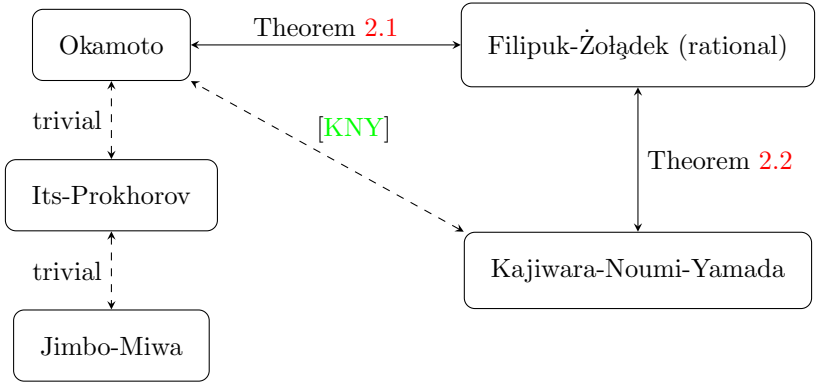
\begin{figure}[ht]
	\begin{center}
	\begin{tikzpicture}[>=stealth,node distance=5mm
	                    ]
	                      
	  \node (KNY) at (4,-1) [box,inner sep=10pt,  align=center]   {Kajiwara-Noumi-Yamada}; 

	  \node (Ok) at (-3,2) [box,inner sep=10pt,  align=center]   {Okamoto};
	  
	  \node (IP) at (-3,0) [box,inner sep=10pt,  align=center]     {Its-Prokhorov};

	  \node  (JM) at (-3,-2) [box,inner sep=10pt,  align=center]    {Jimbo-Miwa};

	  \node (FZ) at (4,2) [box,inner sep=10pt,  align=center]    {Filipuk-\.Zo\l\c{a}dek (rational)};
	  
	  \draw[-latex, <->, dashed] (IP.south)--(JM.north) node[pos=0.5, left] { trivial };
	  \draw[-latex, <->, dashed] (IP.north)--(Ok.south) node[pos=0.5, left] { trivial };
	  \draw[-latex, <->, dashed] (KNY.north west)--(Ok.south east) node[pos=0.5, xshift=10pt, yshift=10pt] { \cite{KNY}  };
	  \draw[-latex, <->] (KNY.north)--(FZ.south) node[pos=0.5, right] { Theorem \ref{Thm:FZtoKNY}  };
	  \draw[-latex, <->] (FZ.west)--(Ok.east) node[pos=0.5, above] { Theorem \ref{Thm:FZtoOk}  };

	\end{tikzpicture}		
	\end{center}
	\caption{Relationships between different Hamiltonian forms of $\pain{III}$}
	\label{fig:allHams}
\end{figure}

\begin{theorem} \label{Thm:FZtoOk}
Let $x=x(t)$ and $y=y(t)$ be solutions of the Filipuk-\.Zo\l\c{a}dek system \eqref{system FZ} and $f=f(t)$ and $g=g(t)$ be solutions of the Okamoto system \eqref{system Ok}. Under the identification of parameters as in \eqref{par Ok}, the following birational transformation is valid:
\begin{equation} \label{change1}
x=f, \quad y=\frac{4f^2 g -2\eta_{\infty }t f^2  -( 2\theta_0+1) f+ 2\eta_0 t}{f^2},
\end{equation} 
with inverse given explicitly by
\begin{equation}
f=x, \quad g=\frac{x^2 y+2\eta_{\infty }t x^2+(2\theta_0 + 1)x -2\eta_0 t}{4x^2}.
\end{equation}
We have the following equality of the 2-forms defining the non-autonomous Hamiltonian structures of the two systems, under the transformation above:
\begin{equation}
dy \wedge dx - d \h^{\FZ} \wedge \, dt = 4 \left( dg \wedge df - d \h^{\Ok} \wedge \, dt \right).
\end{equation}
The Hamiltonian functions themselves are related modulo functions of $t$ by
\begin{equation}
 \h^{\FZ} (x,y,t) = 4 \h^{\Ok}(f,g,t)+ 2 \eta_{\infty} f + \frac{2 \eta_0}{f}.
\end{equation}
\end{theorem}

\begin{theorem} \label{Thm:FZtoKNY}
Let $q=q(s)$ and $p=p(s)$, be solutions of the KNY system \eqref{system KNY} and $x=x(t)$ and $x=y(t)$ be solutions of system \eqref{system FZ} with $\gamma=-\delta=4$. Then with $s=t^2$ the following birational transformation holds:
\begin{equation}
q=t x,\quad p=\frac{2x^2 y+4t x^2+(\beta -2) x-4t}{8 t x^2},
\end{equation}
with parameters related according to 
\begin{equation}
\alpha=-4(a_1-a_2), \quad \beta=4(1-a_1-a_2),
\end{equation} 
or, equivalently, $a_1=(4-\alpha-\beta)/8$, $a_2=(4+\alpha-\beta)/8$.
The inverse transformation is given by
\begin{equation}
x=\frac{q}{\sqrt{s}},  \quad y=\frac{\sqrt{s}(8 p q^2-4q^2  - 2 (1 -2 a_1 -2 a_2 ) q  + 4s)}{2q^2}.
\end{equation}
We again have the 2-forms defining the non-autonomous Hamiltonian structures of the systems equal under the identification above:
\begin{equation}
dy \wedge dx - d \h^{\FZ} \wedge \, dt = 4  \left(dp \wedge dq - d \h^{\KNY} \wedge \,ds \right).
\end{equation}
The Hamiltonian functions themselves are related modulo functions of $t$ by
\begin{equation}
 \h^{\KNY} (q,p,s) = \frac{1}{8 t} \h^{\FZ}(x,y,t)+ \frac{x y }{8 t^2} - \frac{1 }{2t x} . 
\end{equation}
\end{theorem}

\begin{remark}
Note that the exterior derivative in the 2-forms referred to in both of these Theorems is on the 3-dimensional extended phase space so $t$ and $s$ are treated as variables. 
\end{remark}
Both of the above transformations may be verified by direct calculation. 
The transformations and relations between parameters may be obtained using the method presented in \cite{P4paper}, but the geometry of these systems and the calculations establishing this are rather straightforward in this case so we omit them. 

We make remarks at this point concerning the relation of our results to the degenerate cases of the third Painlev\'e equation, in which the geometry of the associated surfaces change. 
The generic case where $\gamma, \delta \neq 0$ has surface type $D_6^{(1)}$ according to Sakai's classification \cite{Sak01}, whereas if $\gamma=0$, $\delta \neq 0$ then the surface type is $D_7^{(1)}$. 
The KNY Hamiltonian in the $D_7^{(1)}$ case is
\begin{equation}  \label{deg1}
K_1=\frac{1}{t}(q^2p^2+q+p t+a_1 q p),
\end{equation}
which leads to the third Painlev\'e equation with parameters 
\begin{equation} \label{deg1params}
\alpha=-8, \quad \beta=-4(a_1-1), \quad \gamma=0, \quad \delta=-4.
\end{equation}
If both $\gamma=0$ and $\delta=0$, then the surface type is $D_8^{(1)}$ and the KNY Hamiltonian further degenerates to
\begin{equation}\label{deg2}
K_2=\frac{1}{t}(q^2 p^2+ q p+q+t/q),
\end{equation}
which gives the third Painlev\'e equation with parameters 
\begin{equation} \label{deg2params}
\alpha=-\beta=-8, \quad \gamma=\delta=0.
\end{equation}
The change in geometry in these cases means that the birational transformations given in Theorem \ref{Thm:FZtoKNY} may cease to work. 
This is indeed the case, but we can make adjustments to obtain transformations that are valid in these degenerate cases.

\begin{proposition}
The $\gamma = 0$, $\delta=-4$ case of the Filipuk-\.Zo\l \c{a}dek Hamiltonian system \eqref{system FZ} is equivalent to the KNY form of the $D_7^{(1)}$ third Painlev\'e equation with Hamiltonian \eqref{deg1}, under the birational transformation
\begin{equation}
q = t x,  \quad p=\frac{x^2 y-2a_1 x+x-2t}{4 t x^2},
\end{equation}
with parameters related according to \eqref{deg1params}.
\end{proposition}

\begin{proposition}
The $\gamma = 0$, $\delta=0$ case of the Filipuk-\.Zo\l \c{a}dek Hamiltonian system \eqref{system FZ} is equivalent to the KNY form of the $D_8^{(1)}$ third Painlev\'e equation with Hamiltonian \eqref{deg2}, under the birational transformation
\begin{equation}
q = t x,  \quad p=\frac{x y-1}{4 t x},
\end{equation}
with parameters related according to \eqref{deg2params}.
\end{proposition}


\bibliographystyle{amsalpha}

\end{document}